\documentclass{aa}
\usepackage{graphics}

\begin{document}

   \thesaurus{01     
              (02.02.1;  
               13.07.3;  
               13.07.1;  
               13.07.2)}  
   \title{The dyadosphere of black holes and 
gamma-ray bursts}


   \author{Giuliano Preparata
          \inst{1}
          \and
          Remo Ruffini\inst{2}
	    \and
          She-Sheng Xue\inst{3}
          }

   \offprints{R. Ruffini}

   \institute{Physics Department, University and INFN-Section of Milan,
Via Celoria 16, 
I-20133 Milan, Italy\\
         \and
             I.C.R.A.-International Center for Relativistic Astrophysics
and
Physics Department, University of Rome ``La Sapienza", I-00185 Rome,
Italy\\
             email: ruffini@icra.it
	\and
     I.C.R.A.-International Center for Relativistic Astrophysics c/o
Physics Department, University of Rome ``La Sapienza", I-00185 Rome,
Italy.
             }

   \date{Received 16 May 1998/Accepted 12 July 1998}

   \maketitle

   \begin{abstract}

  The ``dyadosphere" has been  defined (Ruffini \cite{r2}, Preparata et al. \cite{prx}) as the region outside the horizon  of a
black hole endowed with an electromagnetic field (abbreviated to EMBH
for ``electromagnetic black hole") where the electromagnetic field
exceeds the critical value, predicted by Heisenberg \& Euler (\cite{he}) for $e^+
e^-$ pair production. In a very short time ($\sim O({\hbar\over mc^2})$),
a very large number of pairs is created there. We here give limits on the EMBH
parameters leading to a Dyadosphere for $10M_{\odot}$ and
$10^5M_{\odot}$ EMBH's, and give as well the pair densities as functions of
the radial coordinate. We here assume that the pairs reach thermodynamic
equilibrium with a photon gas and estimate the average
energy per pair as a function of the EMBH mass. These data give the initial conditions for the analysis of an enormous pair-electromagnetic-pulse or ``P.E.M. pulse" which
naturally leads to relativistic expansion. Basic energy requirements for
gamma ray bursts (GRB), including GRB971214 recently observed at $z=3.4$,
can be accounted for by processes occurring in the dyadosphere. In this letter we do not address the problem of forming either the EMBH or the dyadosphere: we establish some inequalities which must be satisfied during their formation process.
   \end{abstract}

%


We {\it assume} that the process of gravitational collapse for a 
core larger than the neutron star critical mass will generally lead to
a black hole characterized by {\it all} three fundamental parameters of
mass-energy $M$, angular momentum $L$, and charge $Q$ (see Ruffini \& Wheeler \cite{rw}). While the first two parameters $M$ and $L$  evolve on very long
time scales, the charge $Q$ will dissipate on much shorter time scales. In
particular we show that if the electromagnetic field of an EMBH is
overcritical in the sense of Heisenberg \& Euler (\cite{he}), it will
dissipate on an extremely short time scale of approximately $10^7$ 
sec.~through a new
process involving the ``dyadosphere" of a black hole. In a very short
time $\sim O({\hbar\over mc^2})$, a very large number of pairs is created
there and reaches thermodynamic equilibrium with a photon gas. In
the ensuing enormous  P.E.M. pulse, a large fraction of the extractable
energy of the EMBH in the sense of Christodoulou-Ruffini (\cite{cr}) will be 
carried away. The P.E.M. pulse will interact with some of the
baryonic matter of the uncollapsed material and  the associated emission
is closely related to the observed properties of GRB sources. The
electromagnetic field of the remnant will further dissipate in the
acceleration of cosmic rays or in the propulsion of jets on much longer
time scales. 

The recent observations of the Beppo-SAX satellite, the discovery of a
very regular afterglow to GRB's, the fact their x-ray flux varies
regularly with time according to precise power laws (Costa  et al.
 \cite{exp}; van Paradijs et al. \cite{p}), and especially the optical and infrared identification of these sources
(Kulkarni et al. \cite{k}, Halpern et al. \cite{ha} and Ramaprakash et al. \cite{ra})
which has established their correct cosmological setting and
determined their formidable energy requirements
have motivated us to reconsider some theoretical results on black hole
astrophysics (see e.g. Ruffini \cite{r1}, Ruffini \cite{r2}) and to develop a detailed
model for a direct confrontation with the observational results. 

The  Christodoulou-Ruffini (\cite{cr}) energy-mass formula
for a Reissner-Nordstrom EMBH gives
\begin{eqnarray}
E^2&=&M^2c^4=\left(M_{\rm ir}c^2 + {Q^2\over2r_+}\right)^2,\label{em}\\
S&=& 4\pi r_+^2=16\pi \left({G^2\over c^4}\right)M^2_{\rm ir},
\label{s}
\end{eqnarray}
with
\begin{equation}
{\sqrt{G}Q\over r_+c^2}\leq 1,
\label{s1}
\end{equation}
where $M_{\rm ir}$ is the irreducible mass, $Q$ the charge and $r_{+}$ the
horizon radius. (We use c.g.s.~units). 
Up to 29$\%$ of the mass-energy of an extreme rotating black hole can be stored as rotational energy and
gedanken experiments have been conceived to extract such energy (see e. g. Ruffini \& Wheeler \cite{rw}). 
In the case of black holes endowed with an electromagnetic field, it follows 
from Eqs.(\ref{em}-\ref{s1}) that up to 50$\%$ of the mass
energy of an extreme EMBH with $Q_{\rm max}=r_+c^2/\sqrt{G}$ can be stored in its
electromagnetic field. It is appropriate to recall that even in the case
of an extreme EMBH the charge to mass ratio is $\sim 10^{18}$ smaller then
the typical charge to mass ratio found in nuclear matter, owing to the
different strength and range of the nuclear and gravitational
interactions. In other words it is enough to
have a difference of one quantum of charge per $10^{18}$ nucleons in
the collapsing matter for an EMBH to be extreme. 

By applying the classic work of
Heisenberg \& Euler (\cite{he}) to EMBH's, as reformulated in a
relativistic-invariant form by Schwinger (\cite{s}),
Damour \& Ruffini (\cite{dr})  showed that a
large fraction of the energy of an EMBH can be extracted by pair
creation.  This energy extraction process only works for
EMBH black holes with $M_{\rm ir}< 10^6M_{\odot}$. They also claimed
that such an energy source might lead to a natural explanation for GRB's and for ultra high energy cosmic rays.

The general considerations presented in Damour \& Ruffini (\cite{dr}) are
correct. However,
that work has an underlying assumption which only surfaces in the very last
formula: that the pairs created in the process of vacuum polarization
are absorbed by the EMBH. That view is now fundamentally modified by the
introduction of the novel concept of the dyadosphere of an EMBH (Ruffini \cite{r2}, Preparata, et al.\cite{prx})
and by the considerations given below.

For reasons of simplicity we use the case of a
nonrotating Reissner-Nordstrom EMBH  to illustrate the basic
gravitational-electrodynamical process. The case of a rotating Kerr-Newmann 
black hole, namely an EMBH with angular momentum, will be considered elsewhere by Damour and Ruffini (in preparation).

\begin{figure}
   \resizebox{\hsize}{8cm}{\includegraphics{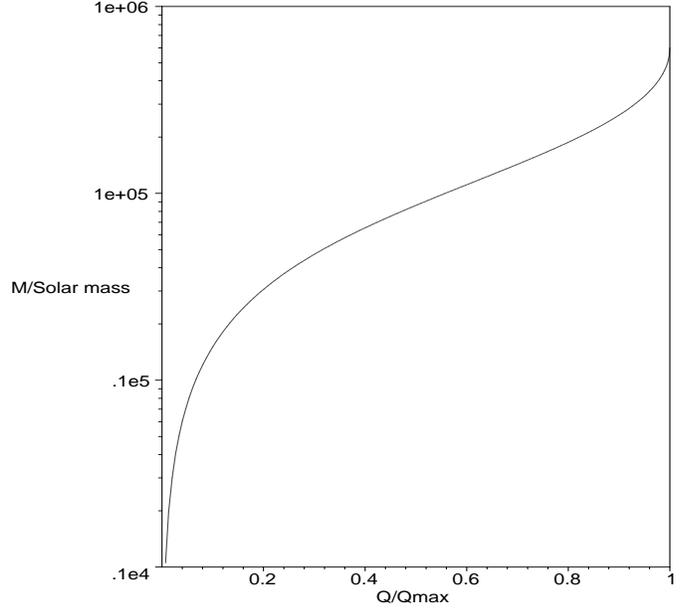}}
      \caption[]{Allowed range of variability of $\mu = {M \over
M_{\odot}}$ and $\xi = {Q \over Q_{\rm max}}$ for which the dyadosphere exists. The allowed region lies below the curve
shown in the figure, see  text.
}
         \label{fig.1}
   \end{figure}

Introducing the dimensionless parameters $\mu={M\over M_{\odot}}>3.2$,
$\xi={Q\over Q_{\rm max}}\le 1$, the horizon radius may be expressed as
\begin{eqnarray}
r_{+}&=&{GM\over c^2}\left[1+\sqrt{1-{Q^2\over GM^2}}\right]\nonumber\\
&=&1.47\cdot 10^5\mu (1+\sqrt{1-\xi^2})\ {\rm cm}.
\label{r+}
\end{eqnarray}
Outside the horizon the electromagnetic field measured by an orthonormal
tetrad at rest at a given radius $r$ in the Boyer-Lindquist
coordinate system (see e.g. Ruffini \cite{r1}) has only one
nonvanishing radial component ${\bf E}= {Q\over r^2}\hat{r}$. 
We can evaluate the radius at which the electric field
reaches the critical value ${\cal
E}_c={m^2c^3\over\hbar e}$ defined by Heisenberg and Euler, where
$m$ is the mass and $e$ the charge of the electron.  
We can then express (Ruffini \cite{r2}, Preparata et al.\cite{prx}) the radius of the dyadosphere in terms of the
Planck charge $q_{\rm p}=(\hbar c)^{1\over2}$ and the Planck mass
$m_{\rm p}=({\hbar c\over G})^{1\over2}$  in the form
\begin{eqnarray}
r_{\rm ds}&=&\left({\hbar\over mc}\right)^{1\over2}\left({GM\over
c^2}\right)^{1\over2} \left({m_{\rm p}\over m}\right)^{1\over2}\left({e\over
q_{\rm p}}\right)^{1\over2}\left({Q\over \sqrt{G}M}\right)^{1\over2}\nonumber\\
&=&1.12\cdot 10^8\sqrt{\mu\xi} \hskip0.1cm {\rm cm},
\label{rc}
\end{eqnarray} 
which clearly illustrates the hybrid gravitational and quantum nature of
this quantity.
The dyadosphere extends over the radial interval
$r_{+}\leq r \leq r_{ds}$. 
It is important to note that the radius of the dyadosphere is maximum
for the extreme value $\xi=1$ and that the dyadosphere exists for EMBH's 
with mass from
the upper limit for neutron stars at $\sim 3.2M_{\odot}$ all
the way up to a maximum mass of $6\cdot 10^5M_{\odot}$;
correspondingly smaller values of the maximum mass are obtained for smaller values of $\xi$, as indicated in Fig.~\ref{fig.1} where we plot the
range of $\xi$ values for which vacuum polarization can occur for selected values of the EMBH mass. The range of variability of $\xi$ is subject to the inequality $\xi_{\rm min}\leq \xi \leq 1$ where  $\xi_{\rm min}$ is implicitly defined by
\begin{equation}
\mu = 6\cdot 10^5{\xi_{\rm min}\over
(1+\sqrt{1-\xi_{\rm min}^2})^2}.
\label{limit}
\end{equation}

The electromagnetic field, which
decreases inversely with the mass, never becomes critical for EMBH's with mass
larger than the maximum value stated above. 

%

The density and energy of pairs created in the
dyadosphere can be modeled (Preparata et al.\cite{prx}) by imagining  the dyadosphere 
to be a sequence of concentric thin shell
spherical capacitors at successive values of the radius $r$,
each of thickness
$\lambda\sim O({\hbar\over mc})$ and charge $\Delta Q(r)$ given by
\begin{equation}
\Delta Q(r)={Q}\big[1-\left({r\over r_{\rm ds}}\right)^2\big].
\label{pair}
\end{equation}
On a very short time scale $O({\hbar\over mc^2})$, the QED vacuum dielectric breakdown produces a number density of pairs $n_{e^+e^-}(r)$ such that $E(r)\approx {\cal E}_{\rm c}$. This approximate equality simply means that the $e^+e^-$ pair creation exponentially decreases when the initial electric field is screened to the critical value. It is important to emphasize that the first layer of thickness $\lambda$ outside the horizon produces a number of pairs sufficient to reduce the charge of the EMBH to $Q_{\rm c}={\cal E}_{\rm c}r_+^2$, its critical value 
in the sense of Heisenberg and Euler. The total number of pairs actually created in the dyadosphere is very much larger than the number captured by the black hole, since it is 
amplified by the factor $(r_{\rm ds}-r_+)/\lambda $.

\begin{figure}
   \resizebox{\hsize}{8cm}{\includegraphics{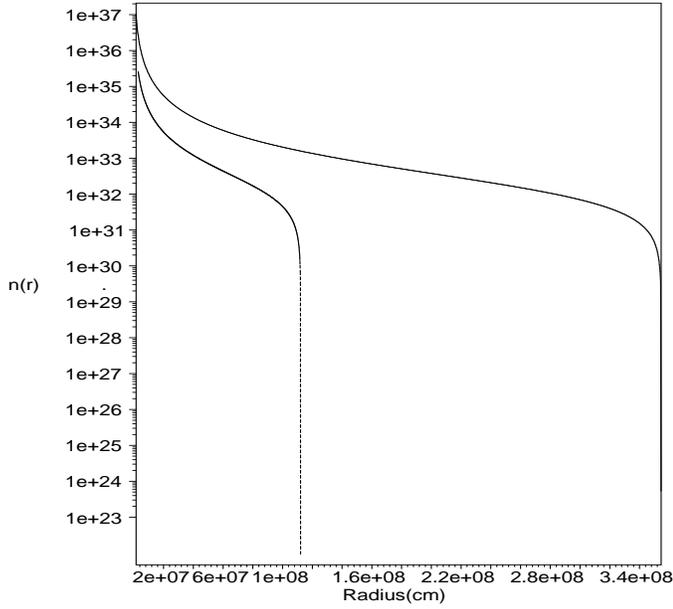}}
      \caption[]{The density $n_{e^+e^-}(r)$
as a function of the radial coordinate for EMBH's of
$10M_\odot$ corresponding to the charge parameter values 
$\xi=1$ (upper curve) and $\xi=0.1$ (lower curve).
}
         \label{fig.2}
   \end{figure}

\begin{figure}
   \resizebox{\hsize}{8cm}{\includegraphics{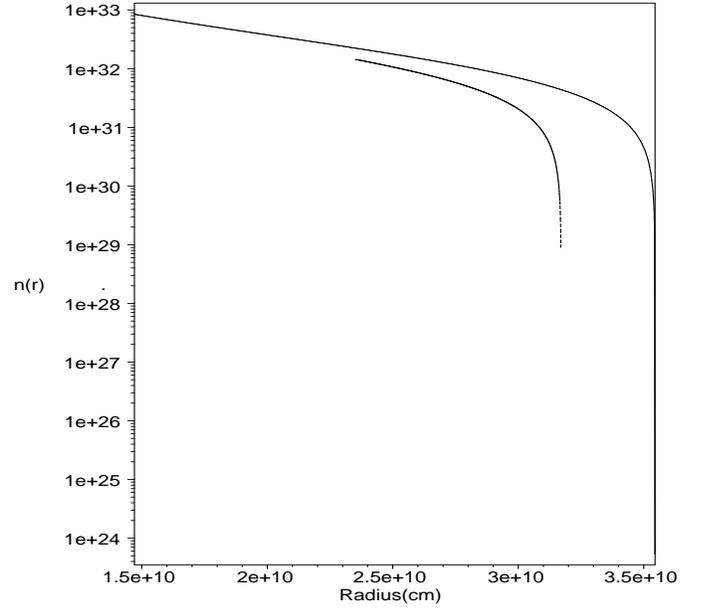}}
      \caption[]{The same as Fig.~\ref{fig.2} for an EMBH of
$10^5M_\odot$ corresponding to the charge parameter values 
$\xi=1$ (upper curve) and $\xi=0.8$ (lower curve).}
         \label{fig.3}
   \end{figure}

	The density of pairs as a function of the radius is then given by
\begin{equation}
n_{e^+e^-}(r)
={1\over 4\pi r^2}\left({Qmc\over e \hbar}\right)
\big[1-\left({r\over r_{ds}}\right)^2\big].
\label{density}
\end{equation}
In Figs.~\ref{fig.2} and \ref{fig.3}, we plot the density of pairs for
$10M_{\odot}$ and $10^5M_{\odot}$ EMBH for selected values of $\xi$. These two values of the mass were chosen to be representative of objects typical of the galactic population or for the nuclei of galaxies compatible with our upper limit of the maximum mass of $6\cdot 10^5M_{\odot}$.

%
%

%
%

We are now in a position to compute the total number of pairs $N_{\rm pair}$ created in the dyadosphere and from a knowledge of
the electrostatic  energy density in each shell,
the energy density of created pairs as a function of the radial
coordinate and the total energy $E^{\rm tot}_{e^+e^-}$ in the pairs.
Finally we can estimate the total energy extracted by the pair creation
process in EMBH's of different masses for selected values of $\xi$ and
compare and contrast these values with the maximum extractable energy
given by the mass formula for black holes (see Eqs.~(\ref{em}) and 
(\ref{s1})). This comparison shows that the  efficiency sharply decreases
as one reaches the maximum value of the EMBH mass permitting vacuum 
polarization, while the efficiency approaches $100\%$
in the low mass limit (Preparata et al. \cite{prx}).

\begin{figure}
   \resizebox{\hsize}{8cm}{\includegraphics{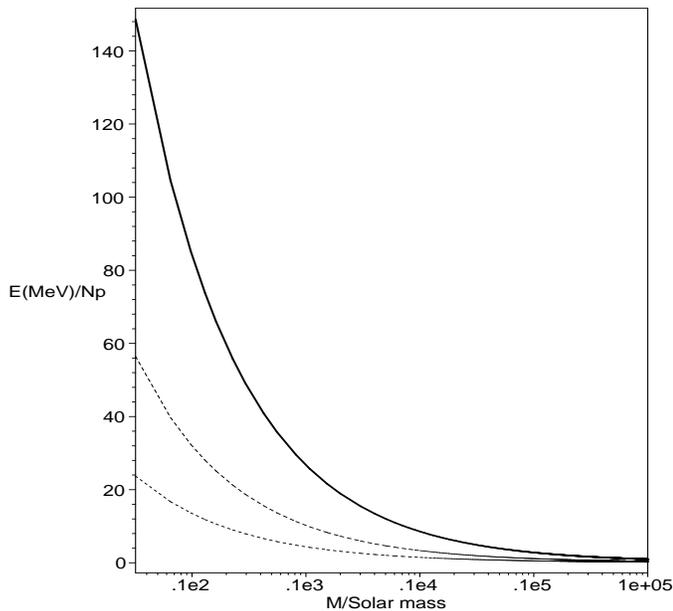}}
      \caption[]{The average energy per pair is shown here as a function of
the EMBH mass in solar mass units 
for $\xi=1$ (solid line), $\xi=0.5$ (dashed line) and
$\xi=0.1$ (dashed and dotted line).}
         \label{fig.4}
   \end{figure}

Due to the very large pair density given by Eq.~(\ref{density}) 
and to the sizes of the
cross-sections for the process $e^+e^-\leftrightarrow \gamma+\gamma$, 
the system is expected to thermalize to a plasma configuration for which
\begin{equation}
N_{e^+}=N_{e^-}=N_{\gamma}=N_{\rm pair}
\label{plasma}
\end{equation}
and reach an average temperature
\begin{equation}
kT_\circ={ E^{\rm tot}_{e^+e^-}\over3N_{\rm pair}\cdot2.7},
\label{t}
\end{equation}
where $k$ is Boltzmann's constant.
The average energy per pair ${ E^{\rm tot}_{e^+e^-}\over N_{\rm pair}}$ is shown as a function
of the EMBH mass for selected values of the charge parameter $\xi$  in
Fig.~\ref{fig.4}.

%

As shown by Ruffini et al.(\cite{rwx}) the further evolution of
this plasma leads to a relativistic expansion, $e^+ e^-$ annihilation and
an enormous pair-electromagnetic-pulse ``P.E.M. pulse". By introducing a
variety of models based on relativistic hydrodynamical
equations, it has been shown that the dyadosphere of
the EMBH reaches relativistic expansion with a relativistic 
gamma factor $100-1000$
within seconds.

If this basic scenario is confirmed by observations, other
fundamental questions should be investigated to understand the
origin of the dyadosphere. Some preliminary work along these lines has
already been done by Wilson (\cite{cw},\cite{jw}) who has shown how relativistic
magneto-hydrodynamical processes
occurring in the accreting material around {\it an already formed black hole}
lead naturally to very effective charge separation and to reaching
the  critical value of the charge given by the limiting value of Eq.~(\ref{s1}) 
on a timescale of $10^2$--$10^3 GM/c^3$.
These studies show the very clear tendency of powerful processes of charge separation and of magnetosphere formation to occur in accretion processes.  They cannot, however, be simply extrapolated to the formation of the dyadosphere.  The time scale of the dyadosphere discharge is of the order of $10^{-19}$ sec (for a detailed discussion including relativistic effects, see Jantzen \& Ruffini \cite{jr}).  Such a time scale is much shorter than the characteristic magnetohydrodynamical time scales.  We expect that the formation of the dyadosphere should only occur during the gravitational collapse itself and {\it in the process of formation} of the EMBH, with the formation of a charge depleted region with an electric field sufficient to polarize the vacuum.
A complementary aspect dealing  with  the extremization and equipartition of the electromagnetic energy in a gravitating rotating object advanced in Ruffini \& Treves (\cite{rt}) should also now be considered again. 
Modeling such a problem seems to be extremely difficult at the present time, also in absence of detailed observations narrowing down the values of the fundamental parameters involved.  Confirmation of the basic predictions of the dyadosphere model by observations of gamma ray bursts and their afterglow would provide motivation and essential information to attack such a difficult problem.

It goes without saying that the Heisenberg \& Euler (\cite{he})
process of vacuum polarization considered here and in Damour \& Ruffini (\cite{dr})
has nothing to do
with the evaporation of black holes considered by Hawking et al.(\cite{h})
either from a qualitative or quantitative point of view. The effective
temperature of black hole evaporation given by Hawking for a
$10M_{\odot}$ black hole is $T\sim 6.2\cdot 10^{-9}K^\circ$, which 
implies a black hole lifetime of $\tau\sim 10^{66}$ years 
and an energy flux of $10^{-24}$ergs/sec, while
a $10^5M_{\odot}$ black hole has a Hawking temperature of $T\sim 6.2\cdot
10^{-13}K^\circ$ and lifetime of $\tau\sim 10^{78}$ years with an energy 
flux of $10^{-32}$ergs/sec!

\end{document}